\documentclass{ws-p8-50x6-00}
\def\lsimeq{\,\,\raise0.14m\hbox{$<$}\kern-0.76em\lower0.28em\hbox
{$\sim$}\,\,}
\begin{document}

\title{LITHIUM-BERYLLIUM-BORON EVOLUTION: FROM MENEGUZZI, AUDOUZE AND REEVES 1971
 UP TO NOW}
\author{M. Cass\'e}
\address{Service d'Astrophysique, DSM/DAPNIA/ Orme des Merisiers, 91191 Saclay CEDEX, France and \\
 Institut d'Astrophysique de Paris, CNRS, 98 bis bd Arago, 75014 Paris, France}
\author{E. Vangioni-Flam}
\address{ Institut d'Astrophysique de Paris, CNRS, 98 bis bd Arago, 75014 Paris, France}
\author{J. Audouze}
\address{ Palais de la D\'ecouverte, avenue F. Roosevelt, 75008 Paris, France and \\
  Institut d'Astrophysique de Paris, CNRS, 98 bis bd Arago 75014 Paris, France}
\maketitle\abstracts{We review the main sources of LiBeB production and show
 that a primary mechanism is at
 work in the early Galaxy involving both ejection and acceleration
 of He, C and O at moderate energy, which by nuclear interaction with
 H and He produce light isotopes. The precise measurement of the Be abundance at [Fe/H] = -3.3
 and of $^6Li$ in halo stars find an explanation in this framework. Thus, 
the preservation of $^6Li$
 in the atmosphere of metal poor stars implied, points toward
 the fact the Spite plateau reflects the primordial value of Li. Consequently,
 it can be used as a baryodensitometer.}

\section{ Basic facts and historical perspective}
 LiBeB are 
exceptional since they are both simple and rare. In the Solar System, their 
abundances amount to Li/H = $2. 10^{-9}$, B/H = $7.6 10^{-10}$  and Be/H = $2.5 10^{-11}$, 
respectively, whereas istopic ratios  of lithium and boron are  $^7Li/^6Li$ = 12.6, 
$^{11}B/^{10}B$ = 4. Indeed they are rare because their nuclei are  fragile. The fact that 
nuclear structures with A = 5 and 8 are violently unstable has limited 
primordial nucleosynthesis by thermonuclear fusion to $^7Li$. Effectively 
Standard BBN is hoplessly ineffective in generating $^6Li$ , $^9Be$, $^{10}B$, $^{11}B$ \cite{E1}.
Stellar nucleosynthesis (quiescent or explosive) produces nuclei from C to U 
but destroy LiBeB, since their burning temperature is especially low.		  

Another mechanism (different from fusion) is thus necessary to yield LiBeB.  This one is 
the break-up of CNO by spallation.
 It is worth noting that among light isotopes, Li, is special, since it owes its 
origin to numerous sources, involving processes as different than BBN 
($^7Li/H$ about $10^{-10}$), nuclear spallation ($<$ $2.10^{-10}$),
 neutrino spallation in SNII ($<$ $2. 
10^{-10}$), and stellar nucleosynthesis through low mass stars and novae (about $10^{-9}$).

The origin of LiBeB is an old problem. Initially, the debate has opposed 
supporters of a Stellar/Circumstellar origin \cite{GFH}
to those of an interstellar origin related to Galactic Cosmic Rays (Reeves 
and collaborators). In a seminal work Meneguzzi, Audouze and Reeves (1971), MAR \cite{MAR}, have given 
credance to the second explanation. They were able to device a consistent 
scenario linking measurements of local abundances, spallation cross sections 
 and the measured GCR flux.
The merit of their work is the clear identification of the production process
i.e.  the Galactic Cosmic Rays / ISM interaction, essentially via fast (p,$\alpha$) impacting on 
CNO. Furthermore, MAR offered a quantitative estimate of the 
present production rate of each isotope, and integrating over time (assuming a 
constant rate) they got local (cumulated) abundances that compare favorably 
with that of $^6Li$ , $^9Be$ and $^{10}B$ in the solar system. 

 The beauty of 
the spallative interpretation is that the abundance hierarchy: B11, B10, Be9 is 
reflected by the production cross sections. This is a 
gratifying  proof that nature follows nuclear physics! 
It is worth emphasizing that alpha induced reaction have lower energy 
threshold than the corresponding proton reactions. This is of importance for 
the postulated low energy component associated to superbubbles (see below).
Despite the success of the MAR approach, some problems remained however, 
concerning the $^7Li/^6Li$ ratio ( the calculated production ratio is $1.2$
 and the observed one in meteorites is 12.6). Thus other 
sources of $^7Li$ were required, namely stars that can produce it through the  
fusion of $^4He$ and $^3He$. Also the $^{11}B/^{10}B$ ratio was discrepant (2.5 against 4 in 
meteorites). Thus an ad-hoc hypothesis was proposed: the production of $^{11}B$ by 
a low energy flux, not observable in the solar cavity.

\section{ Production processes and ratios}

In the 70', LiBeB observations
were limited to the Solar System and closeby stars.  Since the 90's they have 
been continuously pushed down to lower and lower metallicities. The great 
surprise was to find linear correlations between both Be, B and Fe up to 
[Fe/H] =-1, whereas a quadratic relation was expected from the GCR theory.

The production rate of species L is a function of target composition and of the 
projectile characteristics (composition and energy spectrum):
 
 $ dN(L)/dt = N{_T} <\sigma> \phi$

One distinguishes two cases:
If p  and alphas are the projectiles and interstellar CNO, the targets, one deals 
with a Ç Secondary È Production mechanism. 

The fast protons and alphas of the standard GCR react on CNO at rest in the ISM, giving rise to 
LiBeB nuclei. Then:
		
  $dN(L)/dt = N{_{CNO}} <\sigma> \phi{_{\alpha p}}$

Making the reasonable assumption that SNII both produce and eject O (essentially) and 
accelerate GCR (through the shock waves they produce), one gets: 

   $\phi(t)$ $\alpha$ $SN(t)$ $\alpha$ $dO/dt$. The flux decreases as time increases.

  $N{_{CNO}}(t)$ $\alpha$ $SN(t)$ $\alpha$ $O$. The target abundance increases as time increases. Then,

	$dBe/dt$ $\alpha$ $dO/dt.O$ and	$Be/H$  $\alpha (O/H)^2$

Conversely, if C  and O are the  projectiles, and H  and He the targets, the case is different because 
H and He do not evolve significantly in the course of the galactic lifetime, 
their abundance being essentially fixed by the big-bang nuclesosynthesis.  One 
deals, in this case with a Primary Production process.

 $dN'(L)/dt = N{_{H-He}} <\sigma> \phi{_{CO}}$

       $\phi(t)$ $\alpha$ $SN(t)$ $\alpha$ $dO/dt$. Then,

      $dBe/dt$ $\alpha$ $dO/dt$  and $Be/H$ $\alpha$ $O/H$

 This primary process as been astrophysically associated with galactic 
superbubbles (SB) which are interstellar cavities produced by the collective effect of 
SN and WR. In these 
low density regions fresh products of nucleosynthesis (essentially O in our 
case) accumulate and are accelerated by shock waves induced by successive 
supernovae\cite{BY}, \cite{PA}. In 
the following we will call Superbubble Accelerated Particles (SAP)  the fast 
component related to these objects.
Standard GCR and SAP have in common the acceleration agent:  type II 
Supernovae. But, they could differ  by i)  the source composition, SAP being enriched 
w.r.t GCR in O and specifically He (by a factor 3), and this has a strong 
bearing on the $^{6}Li$  problem (see below)
ii) the energy spectrum : the mean energy of GCR is presumably higher than 
that of SAP, 1GeV/n typically, against say 100 MeV/n. 
It is expected that SAP, due to their low energy do not 
escape easily from the SB regions,  and even if they escape, they cannot enter 
the solar cavity due to the repelling effect of the solar wind.  Thus, in this case 
there is some chance that SAP interacting with the dense shells enclosing 
superbubbles could produce C and O gamma ray lines and the LiBe feature (a combination of 
 two broad lines around 450 keV). An existence test of 
this low energy He and  O rich component is the gamma ray emission of local 
OB associations/superbubbles (Cyg OB2, PerOB2, Orion, VelaÉ). This is a good 
prospect for the european INTEGRAL mission (J. Paul, this meeting).  
Above all the two  mechanisms in question  (standard GCR and SAP) differ by the type 
 of production process, secondary in the case of standard GCR,  ($BeB$ $\alpha$ $O^2$)  
and primary in the case of SAP ($BeB$ $\alpha$ $O$).
Before closing this section on physical production processes of light elements, it 
is worth mentionning another primary mechanism, neutrino spallation, which takes place within supernovae. 
It is expected to  generate significant amounts of $^{11}B$ 
and $^{7}Li$ but not Be. However, Nadyozhin (this conference) has warned us that the discussion is 
not closed. Indeed many uncertainties are yet attached to this mechanism.

\section{Empirical O-Fe correlations and evolutionary models}

The [O/Fe] behaviour at low metallicity is a central issue, since the physical correlation is 
between BeB and O, the products and the progenitor, whereas the observable 
one is between BeB and Fe. 
According to the lines analysed (IR triplet, [OI] or permitted OI), the estimates of 
the O abundance differ.

Over a long period, work largely based on the forbidden OI doublet
 observed in giants has established that [O/Fe] levels off below
[Fe/H]= -1 at about 0.5 dex . Consequently, the observed proportionality between
 Be and Fe leads naturally to the proportionality between Be and O. Thus, this implies a clear
 primary origin of BeB.

Subsequently data were released based on UV OH and IR triplet leading 
to [O/Fe] = Ð0.35 [Fe/H]\cite{I} \cite{B}. 
The proportionality between Be and Fe leads in this case to the correlation: 
log(Be/H) proportional to about 1.6 [O/H].
 Thus the slope of the correlation is between 1 
and 2, a quite ambiguous situation, in which a purely secondary origin linked 
to GCR is still possible (\cite{fi1}).

To summarize the situation,  three mechanisms operate: 
i) Standard GCR (secondary) 
ii) SAP (primary)
iii)Neutrino Spallation (primary).

Different models have been proposed to explain the evolution of LiBeB.
i) A pure secondary Standard GCR mode based on variable [O/Fe] \cite{fi1}.
 This one has to confront the following difficulties: is [O/Fe] really 
increasing at low metallicities? Many observers would be reluctant to accept 
this variation. Why O nucleosynthesis should be special compared to 
other alpha elements? Also the energetics of the BeB production in this case is quite 
demanding in the early Galaxy ( Ramaty this conference).

ii) A pure primary component arising from SB assimilated to GCR.  Galactic Cosmic 
Rays in this view becomes primary  \cite{R}. But it seems that their 
source composition cannot be reconciled with the SAP/SN one,  \cite{M}.
 Moreover this kind of model suffers from an underproduction of $^6Li$ in 
the halo due to lack of incident alphas.

iii) An hybrid model giving consecutive roles to SAP (acting in the halo)
 and standard GCR (active in the disk) seems to fulfil the essential observational 
constraints \cite{E2,E3}.

Accurate oxygen abundances of old, metal poor stars are required if one 
desires to reach a satisfactory solution of the halo BeB isssue.
Indeed, a lot of efforts have been put recently on observations and atmosphere 
modelisation and a special session of the IAU at Manchester 2000 has been devoted 
to the O-Fe correlation. The general feeling is that the flat (or slightly increasing)
 [O/Fe] vs [Fe/H] at 
decreasing Fe/H is favoured. The need for a primary component in the early 
Galaxy is thus strenghthened.
 Moreover, the recent observation of Primas et 
al (2000) \cite{PRI} indicating a flatenning of the Be-Fe relation at very low metallicity, 
 reinforces a primary origin of Be through the most massive stars in agreement with the
 prediction of
 Vangioni-Flam et al (1998) \cite{E2}.
 Thus a big-bang origin of Be is not required to explain these data.

\section{ The case of Lithium 6 and 7}

Recently much work has been devoted to the non thermal  production of Li in the early Galaxy, 
especially via the reaction $\alpha + \alpha$. An important condition has to be fulfiled:  lithium 
should not be overproduced at low metallicity. In other words, the Spite plateau should 
not be traversed by the rising Li.   
Recent observations on $^6Li$ in halo stars , amounting to  $^6Li/^7Li$  about  $2$ to $5 
10^{-2}$ (Cayrel and Asplund, this conference) set new constraints on spallative processes 
in the early Galaxy. Indeed GCR (either standard as proposed by Fields and Olive 
or of primary type as proposed by Ramaty et al)
 are not sufficient to explain the observed $^6Li$ 
abundances due to their low alpha content contrary to SAP, 
enriched in alphas \cite{E4}. In the SAP context, it is probable that 
$^6Li$ is almost intact in the atmosphere of halo stars, meaning that $^7Li$, much 
less fragile, has essentially its big-bang value. Accordingly, Li becomes an 
interesting  baryon density indicator (Coc et al, Olive,  this conference ).  

\section{ Conclusion}

Spallation is confirmed as the production process of $^6Li$, Be, B.
The present observational trend is that Fe-O is quite flat at low metallicities indicating a
that primary  production mechanism (SAP) is predominant in halo phase. $^6Li$ 
observations strengthen this assumption since they require an alpha-rich fast component
 in the early Galaxy.
 In the disk phase, standard GCR take increasing 
importance.
The agent of acceleration common to both SAP and GCR  is SNII. The 
favored sources of low energy nuclei (SAP) are massive stars gathered in OB associations
 and exploding in  
superbubbles.

For the future, the needs are the following:

concerning the observational aspect, simultaneous observations O, Fe, Be, B, 
$^6Li$ and $^7Li$ in halo stars would be the most welcome.
 In this context VLT missions are very promising.
A detection of the O* gammay-ray lines and the LiBe feature by INTEGRAL
 (to be launched in 2002) would be a 
 clue to the existence of a low energy He-O rich non thermal component.  
Finally, we would like to draw attention on an intriguing possibility, offering 
a new nucleosynthetic site for LiBeB, namely hypernovae.
Nomoto's velocity and composition profiles of exploding CO cores associated to 
Nadyozhin's analytic treatement of outer layers, allow to determine  the 
number and energy spectrum of C and O particles ejected and the LiBeB they 
produce by impacting on the surrounding medium.  the O + $\alpha$ reactions are 
favoured due to their low energy threshold. A simple rescaling of 
the work of Fields et al (1996) \cite{FIFI} shows a generous production of LiBeB.					

 MAR children and cousins express their aknowledgements to Jean Audouze for his 
major contribution to light element nucleosynthesis and evolution.

\end{document}